\documentclass[11pt]{article}
\begin{document}
\title{Comment on 'Large-scale non-locality in ``doubly special relativity'' with an energy-dependent speed of light'}
\author{Michele Arzano\footnote{\texttt{arzano@physics.unc.edu}}\\{\it Department of Physics and Astronomy,}\\
{\it University of North Carolina,}\\{\it Chapel Hill NC 27599-3255, USA}}
\date{}
\maketitle
\begin{abstract}
We show that some of the recent results reported in gr-qc/0308049 are based on assumptions which are in contrast with
general properties of ``Doubly Special Relativity'' and/or with Planck-scale physics models.
\end{abstract}

\pagenumbering{arabic}
\pagestyle{plain}

\section{Introduction}
In a recent letter \cite{Schutzhold:2003yp}
 Sch{\"u}tzhold and Unruh presented a critical analysis of some possible physical consequences of special relativity theories
with two observer-indipendent scales (also known as ``Doubly Special Relativity'' theories, see e.g. \cite{Amelino-Camelia:2000mn}
\cite{Magueijo:2001cr} \cite{Amelino-Camelia:2002vy}). The purpose of the
present note is to show how the conclusions presented in \cite{Schutzhold:2003yp} rely on assumptions that after 
a careful analysis turn out to be inconsistent. In particular we will focus on the apparent loss of coincidence
for wave packets described in Section V of 
\cite{Schutzhold:2003yp}, the non-local effects described in section IV and we will conclude with a general observation on the questions
raised in Section III. 
\section{Loss of coincidence and field-theoretic example.}
The analysis reported in Section IV of \cite{Schutzhold:2003yp} is based on the definition, given in Section II, of a scalar field of space-time
coordinates and its analogous in energy-momentum space.\\
The main assumption in this field-theoretic model is the possibility of writing a (``doubly special'') Lorentz-transformed
field in space-time using a chain of Fourier transforms. In particular (see equations (4) in \cite{Schutzhold:2003yp})
in order to get the transformed field\footnote{In Section II a model based on a 1+1-dimensional space is introduced,
 later in Section IV the same construction is assumed to be valid in the 3+1-dimensional case.} $\phi'(x,t)$ the authors propose to first Fourier
tranform\footnote{A further assumption made in Section II is that the usual quantum mechanical relations
 $ip\rightarrow\hbar\frac{\partial}{\partial x}$, etc. hold. In Planckian regimes this seems to be unlikely since
it is well known \cite{Maggiore:rv} that quantum-gravitational effects and the presence of a minimal length 
would give us a ``Generalized Uncertainty Principle'' instead of the usual Heisenberg uncertainty relation.}
$\phi(x,t)$ to some $\tilde{\phi}(E,p)$ then to
make a non-linear relativistic transformation in energy-momentum obtaining $\tilde{\phi}(E',p')$ and finally
Fourier transform back this field to coordinate space to get the space-time rotated $\phi'(x,t)$
\footnote{A first attempt in the construction of a consistent field theory formalism
in the special case of $\kappa$-Minkowski space-time has already been made e.g. in \cite{Amelino-Camelia:2001fd}. 
There some first physical consequences for a
scalar field with interaction were derived.}
.\\
The validity of the previous construction can be objected if one observes that
 the definition of Fourier transformation and its inverse in the context of the modified relativity theories
under consideration should be handled with care. 
In fact both in 
energy-momentum space and in coordinate space-time the integration process could turn out to be non-trivial.
 Although the space-time structure of DSR theories is not yet completely clear, it has been observed \cite{Amelino-Camelia:2002vy} 
\cite{Kowalski-Glikman:2002jr} that
a special feature of the new theories
might be the non-commmutativity of space-time coordinates
 with an associated Hopf algebra structure of symmetries \cite{Agostini:2003vg}. These more complex structures possess  
 a non-trivial integral calculus in which a prescription of ``normal ordering'' of the non-commuting coordinates 
is essential. For example in the case of a $\kappa$-Minkowski non-commutative space-time (which, with its (Hopf) algebra of symmetries,
is one of the most studied examples of DSR-like space-times) the integral of a function over the non-commuting coordinates is 

\begin{displaymath}
\int f(x)\,\, d^4x=\int_{S,R}d^4x\,\, \Omega_{S,R}(f)
\end{displaymath}
where the subscripts $S,R$ indicate two different kind of normal ordering of the coordinates (time ``symmetrized'' and time to the right)
which for field theoretic purposes turn out to be equivalent \cite{Agostini:2003vg}.
We can see that in equation (8) of \cite{Schutzhold:2003yp} the normal ordering of the non-commuting coordinates in the exponential would give
some extra (non-trivial, i.e. depending on space coordinates) factors which would compromise the validity of equation (10) and thus
the whole set of conclusions derived from it.\\
 Moreover the general structure of the yet-to-be-built DSR space-time
(not necessarily a type of space-time non-commutativity) might require the presence of an integration measure 
invariant under the action of a new structure of symmetries 
which could introduce another factor in equation (10) again affecting the final result.\\ 
But even if we restrict our considerations to the energy-momentum space,
without specifying any property of the DSR space-time, we have to face a similar problem. 
In this case, in fact, it is obvious that in order to guarantee the 
invariance of the integration measure under the non-linear DSR transformations 
(which can be viewed as a dimensionful deformation of ordinary
Lorentz maps)
a non-trivial extra factor should be present in equation (10).
\\Any one of the arguments above is sufficient to conclude that at the end of the back and forth Fourier-Lorentz
transforming process the equalities in equations (10) of \cite{Schutzhold:2003yp} do not hold anymore.
This means that at the present stage of knowledge of the properties of DSR theories 
the conclusions about the loss of coincidence of wave packets in Section V of \cite{Schutzhold:2003yp} are not valid.
\\Furthermore we could recall that one of the key aspects of a DSR framework is the presence of an upper/lower
bound for energies/lengths. This physical feature turns out to be rather embarassing when we try to define a Fourier 
transfrom integrating over all the range of energies/lengths. 
\section{Energy-dependent speed of light and non-locality}
A questionable assumption is also at the base of the conclusions reached in Section IV of \cite{Schutzhold:2003yp}. In that Section it is argued that
the varying speed of light scenario emerging from the non-linear structure of DSR leads to non-local effects on the large scale behaviour
of Planckian particles. Following \cite{Amelino-Camelia:2002tc} the authors consider the  speed of massless particles
 given by the usual Hamiltonian relation $v=\frac{E}{p}$ which implies, for a general non-linear DSR dispersion relation $E(p)$, an 
energy dependence of the speed of light. 
Their claim is that if we can follow the trajectory of a Planckian (with a Planckian energy) particle over a macroscopic distance 
and time (larger than Planck scale and Planck time) then we would get non-local effects.
The possibilty of localizing such a particle is required by the need for an operational definition of velocity for
our particle-picture based theory.\\
The basic problem here is that an operational definition of velocity, as the one given by the authors of \cite{Schutzhold:2003yp}, 
turns out to be problematic in a Planckian regime. The localization procedure itself would be dramatically affected in
this extreme energy
regime. In fact in order to localize with a proper accuracy a Planckian particle
one needs Planckian particles which are the only ones able to give the resolution we need. These ``detecting'' particles will behave
according to the unusual schemes of our theory. In particular (besides the obvious difficulties due to their ``varying speed'' properties)
 their interactions will be rather complicated and the actual
status of the theory can make few predictions on what would eventually happen from  a dynamical point of view. Thus whatever kind of 
detecting device we may try to use it will never 
be accurate enough to give a real meaning to the concept of 
``macroscopic trajectory of a Planckian particle'' 
.\\ 
In spite of the fact that detections in ``laboratory'' based experiment could be problematic,
the varying speed of light aspects of some DSR theories
can indeed predict interesting (and detectable) physical effects like the in-vacuo dispersion
for gamma-ray bursts travelling cosmological distances (see for example \cite{Amelino-Camelia:2003dg}).

\section{Observation on the energy of composite systems}
For a final comment we will focus on the problem of the energy of composite systems discussed in section III of \cite{Schutzhold:2003yp} 
\footnote{This issue has already been debated in the literature as the authors point out in the final ``note added'' in \cite{Schutzhold:2003yp}.}.
\\It is obvious that macroscopic bodies have masses well above the Planck energy and do not show any of the DSR-like properties. 
On the other hand the fact that the Compton
wavelength of an elementary particle has to be larger than the Planck length implies that the Planck energy sets an upper limit
for the mass of any elementary particle \cite{Schiller:1996fw}.\\ Our new relativistic tranformations involving a maximum energy are then
valid only for elementary particles in their quantum realm. In particular the transformations of DSR theories will describe the basic
relativistic symmetries of a ``one (elementary) particle system''.
As a consequence we see that the laws of composition of energy and momentum for 
composite systems need to be derived from a proper ``multiparticle'' theory. Such a theory should be  the analogous of a`` quantum field theory'' 
formulated in the proper framework of ``generalized'' Lorentz invariance. Only when such a theory has been proposed we can make
predictions  on processes involving several elementary
particles and their interactions and in which each individual energy will be bounded by the Planck energy.

\section*{Acknowledgments}
I would like to thank Y. Jack Ng for valuable comments and suggestions. 
I also thank Stephen Lau for useful discussions.

\baselineskip 12pt plus .5pt minus .5pt

\end{document}